\documentclass[conference]{IEEEtran}
\IEEEoverridecommandlockouts
\usepackage{ulem}
\newcommand{\ra}[1]{\renewcommand{\arraystretch}{#1}}
\usepackage{color}
\usepackage{booktabs} % For formal tables
\usepackage{graphics}
\usepackage{graphicx} 
\usepackage[T1]{fontenc}
\usepackage[utf8]{inputenc}
\usepackage{subfigure}
\usepackage{balance}
\usepackage{amsmath}
\usepackage{colortbl}
\usepackage{tabularx}
\usepackage{xspace}
\usepackage{graphicx}    % For importing graphics
\usepackage{url} 
\usepackage[square,sort,comma,numbers]{natbib}
\usepackage{threeparttable}     
\usepackage{multirow}
\usepackage{caption}
\def\BibTeX{{\rm B\kern-.05em{\sc i\kern-.025em b}\kern-.08em
    T\kern-.1667em\lower.7ex\hbox{E}\kern-.125emX}}
\begin{document}

%\title{Intelligent Reflective Surface Assisted UAV Communication }
\title{Enhancing Cellular Communications for UAVs \\
via Intelligent Reflective Surface }

\author{\IEEEauthorblockN{Dong Ma}
\IEEEauthorblockA{\textit{School of Computer Science \& Engineering}\\
\textit{University of New South Wales}\\
Sydney, Australia \\
dong.ma1@unsw.edu.au}
\and
\IEEEauthorblockN{Ming Ding}
\IEEEauthorblockA{
\textit{Data61, CSIRO}\\
Sydney, Australia \\
ming.ding@data61.csiro.au}
\and
\IEEEauthorblockN{Mahbub Hassan}
\IEEEauthorblockA{\textit{School of Computer Science \& Engineering}\\
\textit{University of New South Wales}\\
Sydney, Australia  \\
mahbub.hassan@unsw.edu.au}
}

\maketitle

\begin{abstract}
Intelligent reflective surfaces (IRSs) capable of reconfiguring their electromagnetic absorption and reflection properties in real-time are offering unprecedented opportunities to enhance wireless communication experience in challenging environments. In this paper, we analyze the potential of IRS in enhancing cellular communications for UAVs, which currently suffers from poor signal strength due to the down-tilt of base station antennas optimized to serve ground users. We consider deployment of IRS on building walls, which can be remotely configured by cellular base stations to coherently direct the reflected radio waves towards specific UAVs in order to increase their received signal strengths. Using the recently released 3GPP ground-to-air channel models, we analyze the signal gains at UAVs due to the IRS deployments as a function of UAV height as well as various IRS parameters including size, altitude, and distance from base station. Our analysis suggests that even with a small IRS, we can achieve significant signal gain for UAVs flying above the cellular base station. We also find that the maximum gain can be achieved by optimizing the location of IRS including its altitude and distance to BS.
\end{abstract}


\section{Introduction}
Intelligent reflective surface (IRS)~\cite{li2017electromagnetic}, also known as meta-surface, is a recently realized artificial material whose electromagnetic properties, such as absorption, reflection, refraction, and phase can be configured electronically in real time. These surfaces can be manufactured at low cost and as such they can be deployed universally providing an unprecedented opportunity to control the wireless multipath environment both indoor and outdoor. IRS therefore opens up a completely new direction in wireless communications research where the focus shifts from combating the multipath to designing it. Indeed, many researchers have recently confirmed that IRS-assisted solutions can significantly improve the capacity, coverage and energy efficiency of existing mobile networks~\cite{huang2018energy,wu2019beamforming,guo2019weighted,yu2019miso,tan2018enabling}. 

In this paper, we explore the potential of IRS to solve a new problem facing the cellular networks. With the proliferation of industrial and civilian use of unmanned aerial vehicles (UAVs), a.k.a drones, there is a growing need and interest to provide cellular connectivity to UAVs. However, because the base station antennas are down tilted, i.e., their main lobes pointing to the ground optimizing coverage for ground users, UAVs flying above the base stations are only supported through the side lobes. Indeed, recent 3GPP studies have confirmed that UAVs receive poor signals from existing terrestrial base stations, and hence supporting aerial users will require further research and development~\cite{fotouhi2019survey}. The question we seek to answer in this paper is: to what extend IRS can help improve cellular reception for UAVs?       

To answer this question, we analyze a future smart city scenario where building facades are instrumented with IRS, which can be remotely programmed or configured by base stations in real-time. Specifically, we assume that, under appropriate control signals, the IRS is capable of reflecting impinging base station signals toward the target UAV with appropriate phase shifts so the reflected waves combine constructively with original waves from the base station to produce a strong signal at the UAV. We then analyze the signal gain at the UAV as a function of UAV height as well as various IRS parameters including its size, altitude, and distance from base station. Our analysis reveals that a small IRS patch on a building facade can potentially provide a 21dB gain for UAVs.  

The rest of the paper is structured as follows. Section \ref{sec:irs} provides a background on IRS. We explain the system model of the proposed IRS-assisted cellular UAV communications in Section \ref{s:system_model}. Simulation experiments and results are discussed in Section \ref{s:simulation}. We review related work in Section \ref{related} before concluding the paper in Section \ref{s:conclusion}.

% The communication enhancement for IRS-assisted communication comes from two folds. First, due to the down-tilt pattern of the cellular base station, IRS acts as a ground user to gather the energy and re-emits it directly to UAV. Second, with the capability to control the phase of reflected signals, IRS enables a coherent combination of signals at UAV. 

\section{IRS Background}  
\label{sec:irs}
\begin{figure}[t]
	\centering
	\centering
		\includegraphics[scale=0.9]{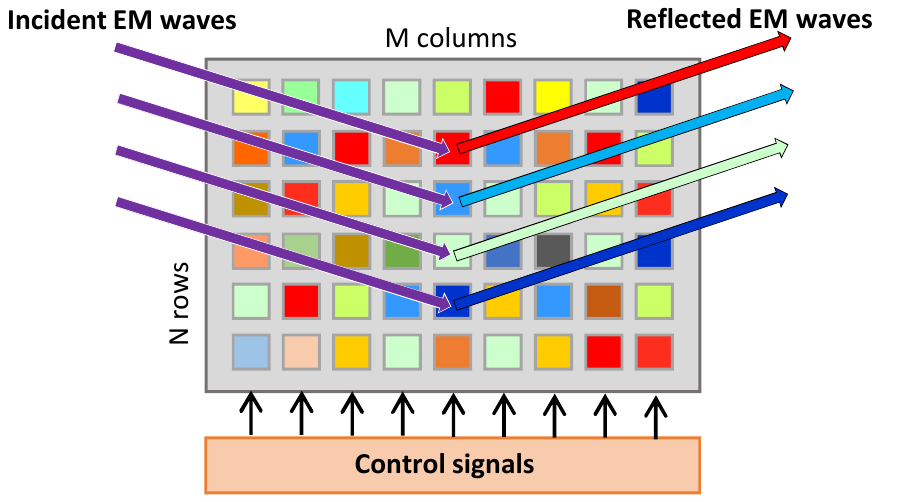}
	\caption{Schematic of intelligent reflective radio surface (IRS).}
	\label{fig:irs}
\end{figure}

IRS is a novel reflective radio technology that is attracting growing attention in recent years~\cite{liaskos2018realizing,zhang2018space}. As shown in Figure~\ref{fig:irs}, IRS presents a two-dimensional artificial structure with a large number of passive reflective elements, whose electromagnetic characteristics, e.g., scattering, reflection, and refraction, can be controlled independently and electronically in real-time by applying different control signals. As a result, the phase and amplitude of impinged electromagnetic waves can be reconfigured and reflected in a software-defined way.  Moreover, the direction of the reflected signal can be controlled precisely to a desired receiver~\cite{diaz2017metasurfaces,diaz2017generalized}, enabling a completely programmable radio environment. Such unprecedented capability to program the radio space promises huge potential in wireless communications. For example, by coherently reflecting the signal from a transmitter to the receiver, IRS can improve the received signal-to-noise ratio (SNR) thereby enhancing the coverage and throughput of wireless networks~\cite{guo2019weighted}. Similarly, IRS can be used for cancelling or suppressing harmful wireless interference by controlling the propagation of all radio waves in a given space of interest~\cite{tan2016increasing}.

There are three other characteristics of IRS that are crucial to the realization of the above mentioned utilities and practical deployment at scale. First, IRS is made of passive elements like printed dipoles that are both energy-efficient as well as cost-effective~\cite{wu2018intelligent}. Second, IRS can be fabricated with high density~\cite{tang2019wireless,tang2019programmable}. For example, an experimental prototype presented in~\cite{tang2019wireless} has an area of $176 \times 252.8 mm^2$ with $8 \times 16$ unit cells, achieving a density of 2876 elements per square meter.
Third, IRS can be electronically controlled with a fast switching rate between different states, which would allow reconfiguration of reflected waves in real-time. For example, the authors of~\cite{tang2019wireless} have achieved 1.25M per second switching rate for IRS, which would allow changing the phase of a reflected multi-path wave within 800ns.

Next, let us examine how IRS elements reconfigure the amplitude and phase of the reflected EM waves. Let $M$ and $N$ respectively denote the number of rows and columns of reflective elements in an IRS. Assume there is an electromagnetic (EM) wave $E_{n,m}$ that impinges on an element $(n,m)$, where $1 \leq n\leq N$ and $1 \leq m\leq M$. The amplitude and phase shift reflection coefficient of element $(n,m)$ are denoted by $A_{n,m}$ and $e^{j \varphi_{n,m}}$, respectively. Then the reflected EM wave $\widetilde{E}_{n,m}$ from this element can be expressed as~\cite{tang2019wireless}

\begin{equation}
\widetilde{E}_{n,m} =A_{n,m} \cdot{e^{j \varphi_{n,m}}}  \cdot{E_{n,m}}
\end{equation} 

At a specific location $\mathbf{s}$ near the IRS, the overall reflected EM wave $\widetilde{E}(\mathbf{s})$ is the superposition of waves reflected by all IRS elements, which also depends on the wireless channel $h_{n,m}(\mathbf{s})$ between each element and $\mathbf{s}$. $\widetilde{E}(\mathbf{s})$ is obtained as~\cite{tang2019wireless}

\begin{equation}
\begin{aligned}
\widetilde{E}(\mathbf{s}) &= \sum_{n=1}^N \sum_{m=1}^M h_{n,m}(\mathbf{s})\cdot \widetilde{E}_{n,m} \\ 
&=  \sum_{n=1}^N \sum_{m=1}^M h_{n,m}(\mathbf{s})\cdot A_{n,m} \cdot{e^{j}} %\varphi_{n,m}}}  \cdot{E_{n,m}}
\end{aligned}
\end{equation}

\section{System Model}
\label{s:system_model}
In this section, we present the system model for the proposed IRS-assisted cellular UAV communications.  

\begin{figure}[t]
	\centering
	\includegraphics[scale=0.45]{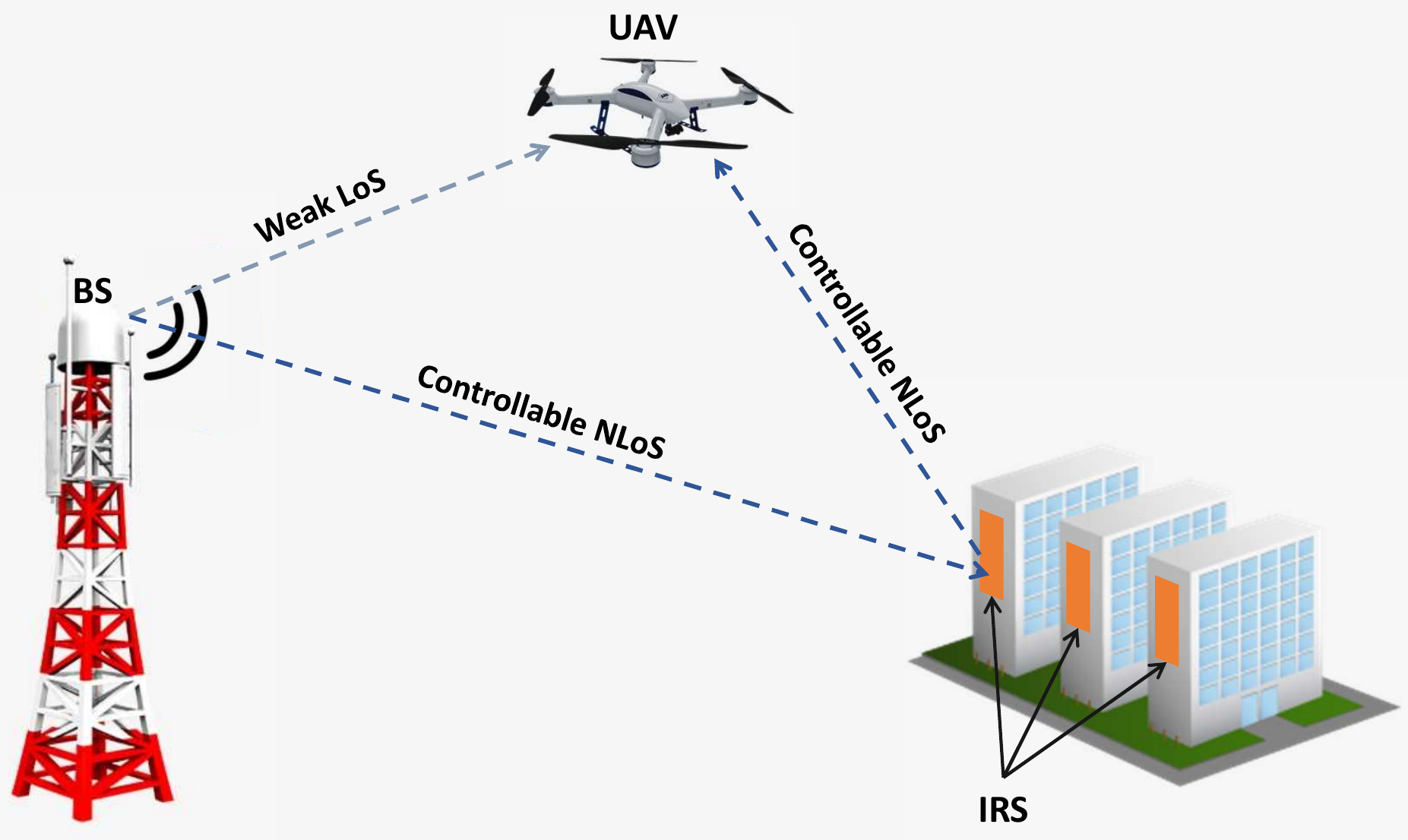}	
	\caption{System model for IRS-assisted cellular UAV communications.}
	\label{fig:system_model}
\end{figure}

\subsection{IRS-assisted UAV Communication System}
As illustrated in Figure~\ref{fig:system_model},
we consider a point-to-point communications system, 
where the IRS is employed to facilitate the downlink (DL) cellular communication. 
The BS is equipped with a single antenna that exhibits a down-tilt antenna pattern. 
The IRS comprises of $K = M \times N$ elements,
where each element can be programmed and reconfigured dynamically to control the direction and phase of the reflected signal.
Specifically, 
once the location of the UAV is available, 
the controller can calculate the direction and phase shift of the reflected signal from each IRS element so that these signals can be coherently combined at UAV. 
We assume perfect and continuous control of both the direction and phase shift. 
The UAV receives signal directly from the line-of-sight (LoS) path as well as the non-line-of-sight (NLoS) paths reflected by IRS elements. 
By aligning the phase of the individual reflected signals to that of the LoS path, these signals are constructively combined at the UAV, 
enhancing the received signal power.

In each scheduling time slot, 
the BS transmits a signal $X$ to the UAV. 
The received signal $Y$ at the UAV can be written as
\begin{equation}
    Y=h_0X + \sum_{i=1}^{K}h_iX + u,
\end{equation}
where $h_0X$ represents the signal received from the LoS path, 
$\sum_{i=1}^{K}h_iX$ is the superposition of signals reflected by the $K$ IRS elements, 
and $u$ denotes additive white Gaussian noise (AWGN) at UAV with zero mean and a variance of $\sigma^2$. 
With the capability to control the phase of the reflected signal using IRS, 
the received signals can be constructively combined to achieve higher received signal power. 

\subsection{Antenna Model}
Typically, 
the LoS signal will dominate the received signal. 
However, 
a BS in the current cellular network is equipped with down-tilt antenna to serve the ground users. 
As a result, 
when UAV is flying higher than the BS, 
the power dissipated through LoS is much weaker than that through NLoS. 
Based on the 3GPP TR 36.814~\cite{trseries}, 
the vertical antenna gain at angle $\theta$ can be modeled as
\begin{equation}
    A(\theta)=-min[12(\frac{\theta-\theta_{etilt}}{\theta_{3dB}})^2, SLA],
\end{equation}
where $\theta_{etilt}$ is the electrical antenna downtilt and its value is set to 15 degree in 3GPP standard. 
$\theta_{3dB}$ equals to 10 and $SLA=20dB$. 
As a result, 
given transmission power of $P_T$ at BS, 
the effective transmission power at angle $\theta$ is
\begin{equation}
    P_T(\theta)=P_T+A(\theta).
\end{equation}

\begin{table}[t]
           \small 
           \centering
           \setlength{\abovecaptionskip}{0pt}
           \setlength{\belowcaptionskip}{10pt}
           \caption{System Parameters.}
           \vspace{0.1in}
           \label{tab:parameters}
           \ra{1.3}
           \begin{tabular}{@{}|l|r|r|@{}}
           \hline
           \textbf{Parameter}&  \textbf{Symbol} &\textbf{(Default) Value}  \\ \hline
            Carrier frequency & $f$ & 2GHz  \\ \hline
            BS transmission power& $P_T$ & 46 dBm \\ \hline
            BS antenna down-tilt angle& $\theta_{etilt}$ & $15^{\circ}$ \\ \hline
            Reflection power loss of IRS & $PL_{IRS}$ & 1dB   \\ \hline
            Reflection power loss of wall & $PL_{wall}$ & 10dB   \\ \hline
            Height of BS & $H_{BS}$ & 25m   \\ \hline
            Height of UAV &$H_{UAV}$ & 50m  \\ \hline
            Height of IRS center &$H_{IRS}$ & 10m  \\ \hline
            Number of IRS elements & $K$ & 100 \\ \hline
            Distance between BS and IRS & $L$ & 50m \\ \hline
\end{tabular}
\end{table}

\subsection{Path Loss Model}
We consider the path loss model for urban macro scenario as presented in 3GPP TR 38.901~\cite{trseries}. 
The path loss $PL_{LoS}$ for the LoS can be calculated with
\begin{equation}
    PL_{LoS} = 28 + 22log_{10}(d) + 20log_{10}(f),
\end{equation}
where $d$ is the distance between the BS and UAV and $f$ is the carrier frequency in GHz.

The path loss $PL_{NLoS}$ for the NLoS is calculated using
\begin{equation}
    PL_{NLoS} = 
\left\{
             \begin{array}{lr}
             max(PL_{LoS}, PL_0), H_{UAV} < 22.5m &  \\
             PL_1, H_{UAV} > 22.5m &  
             \end{array}
\right.
\end{equation}
where $H_{UAV}$ is the height of UAV in meters, and
\begin{equation}
    PL_0=13.54+39.08log_{10}(d) + 20log_{10}(f) - 0.6(H_{UAV}-1.5),
\end{equation}
\begin{equation}
    PL_1=-17.5+(46-7log_{10}(H_{UAV}))log_{10}(d) \\ +20log_{10}(\frac{40 \pi f}{3}).
\end{equation}
The above path loss model considers power loss due to sphere radiation as well as distance-related loss. 

Specifically, 
the received signal from each IRS element $k, (k=1,...,K)$ experiences two paths, 
i.e., BS to element $k$ and element $k$ to UAV with distance $d_1(k)$ and $d_2(k)$, respectively. 
The path loss from BS to IRS can be calculated with
\begin{equation}
    PL_{BS-k}=PL_{NLoS}(d_1(k)).
\end{equation}
For the IRS to UAV path, 
since the received power at IRS is directly reflected to UAV, 
we consider the distance-related component only, i.e.,
\begin{equation}
    PL_{k-UAV}=PL_{NLoS}(d_1(k)+d_2(k))-PL_{NLoS}(d_1(k)).
\end{equation}
In addition, 
when the radio wave impinges on the IRS, 
we assume there is $PL_{IRS}$ dB power loss.
Thus, 
the received power from $k$th IRS element can be expressed as
\begin{equation}
    P_R(k) = P_T(\theta_k) - PL_{BS-k} - PL_{IRS} - PL_{k-UAV},
\end{equation}
where $\theta_k$ is the angle between the path from BS to element $k$ and the BS broadside. The corresponding channel coefficient $h(k)$ is written as
\begin{equation}
    h(k) = \sqrt{P_R(k)}e^{-j \varphi_k},
\end{equation}
where $\varphi_k$ is the phase of the received signal from element $k$. 

Similarly, 
the received power and channel coefficient of the LoS path are
\begin{equation}
    P_R(LoS) = P_T(\theta_0) - PL_{LoS},
\end{equation}
\begin{equation}
    h(0) = \sqrt{P_R(LoS)}e^{-j \varphi_0},
\end{equation}
where $\theta_0$ is the angle between the LoS path and the BS broadside, 
and $\varphi_0$ is the phase of the LoS signal.

Accordingly, 
the received signal amplitude at UAV is given by
\begin{equation}
    \gamma = |h(0)+\sum_{i=1}^{K}h(k)|.
\end{equation}

 \begin{figure}[t]
	\centering
	\centering
		\includegraphics[scale=0.63]{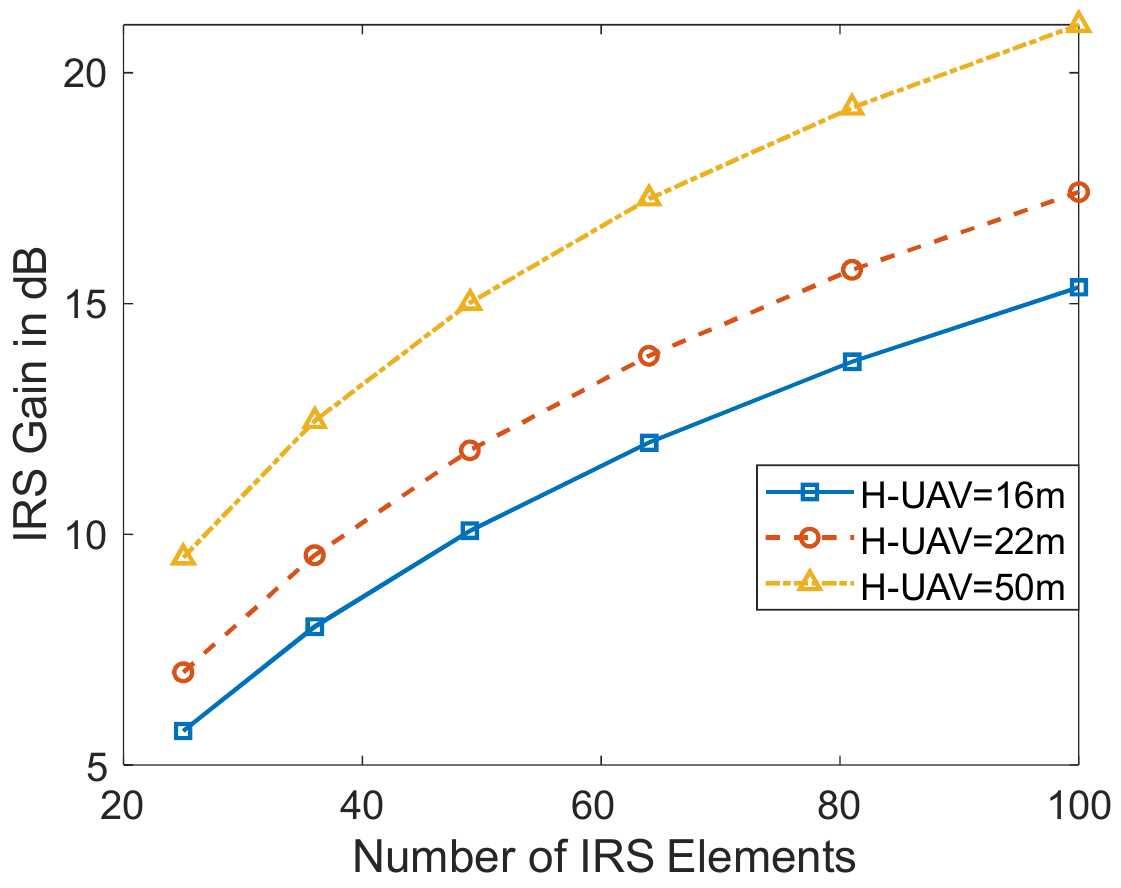}
	\caption{IRS gain vs. number of IRS elements.}
	\label{fig:irs_numer}
\end{figure}

\begin{figure*}[t]
	\centering
	\subfigure[]{
		\includegraphics[scale=0.63]{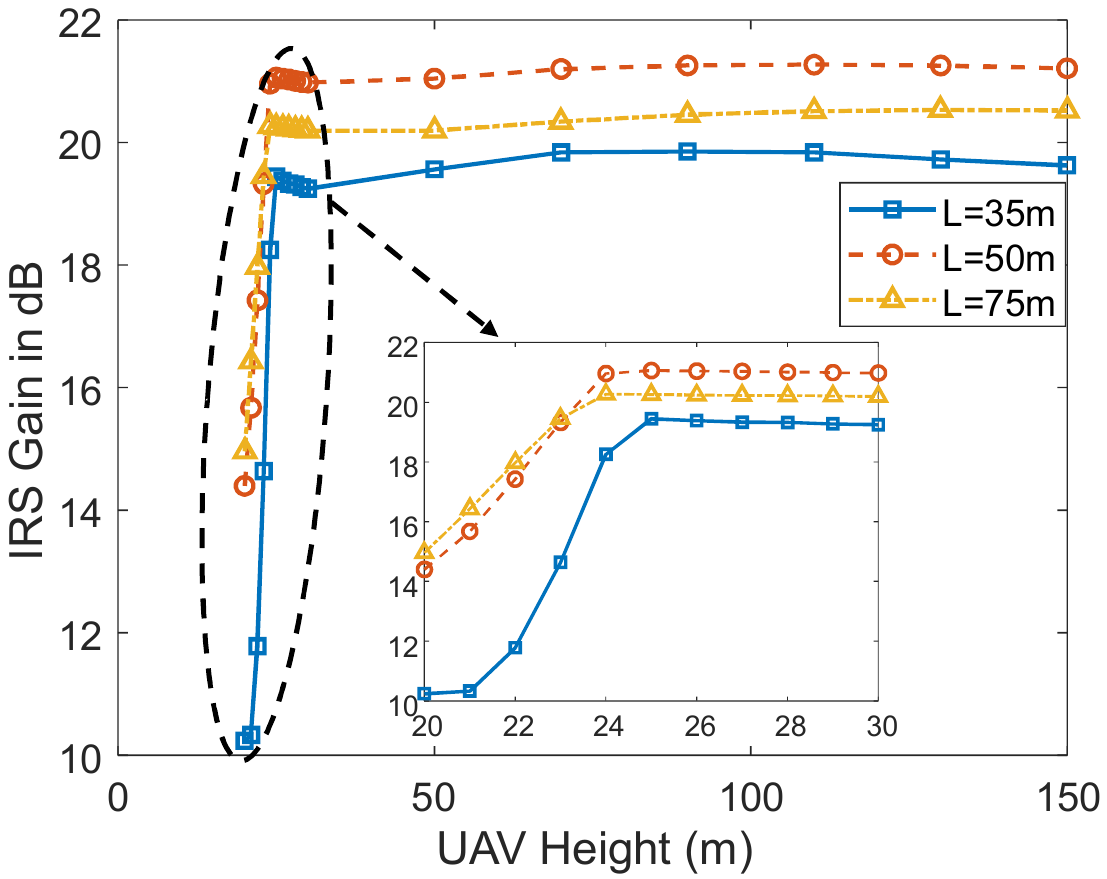}}
		\hspace{0.5in}
	\subfigure[]{
		\includegraphics[scale=0.63]{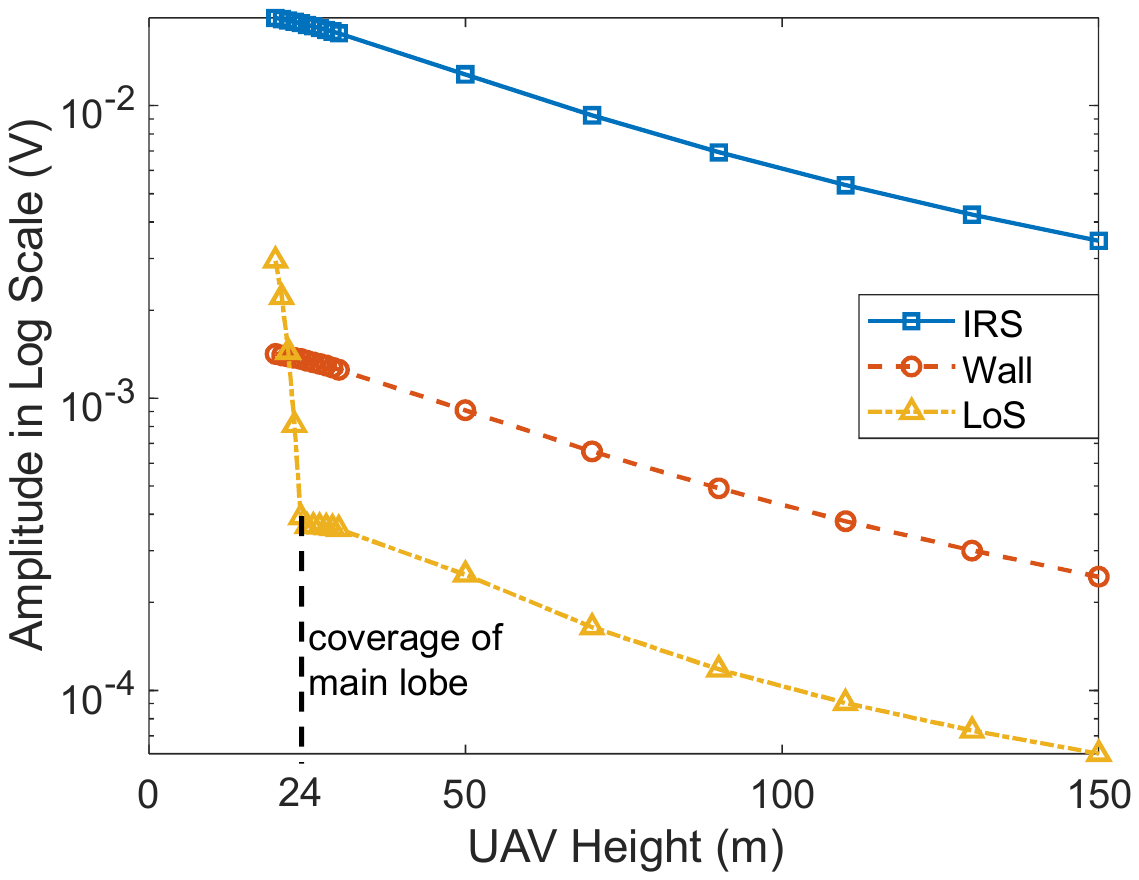}}
	\caption{(a) IRS gain vs. $H_{UAV}$, (b) received signal amplitude from IRS, wall and LoS links.}
	\label{fig:2d_distance}
\end{figure*}

\section{Simulation Results}
\label{s:simulation}
We investigate the performance of IRS-assisted UAV communication via simulation and benchmark with the non-IRS scenario, 
i.e., the wall of a building scatters the incident EM waves to arbitrary directions.
Specifically, 
we define \textit{IRS gain} as the ratio of received signal powers with and without IRS, i.e., $\gamma_{IRS} / \gamma_{non-IRS} $, 
and use it as the metric to evaluate the performance of IRS-assited UAV communication.
For the non-IRS scenario, 
following the 3GPP TR25.996~\cite{trseries}, 
we assume that UAV can receive 20 rays reflected from the same area of deploying IRS on the wall.
Note that the phase of these rays depends on the distance of the propagation path only as no IRS is used to reconfigure the phase shift.
Regarding the power loss due to reflection, 
we assume 1dB loss ($PL_{IRS}$) for IRS due to its capability to control the reflection amplitude coefficient and 10dB loss for the wall. 
We consider a square IRS with K elements, 
i.e., $\sqrt{K}$ rows and $\sqrt{K}$ columns,
where the center of adjacent elements is separated by 2cm.  

The carrier frequency $f$, 
BS height $H_{BS}$, 
and transmission power $P_T$ are set to 2GHz, 25m, and 46dBm, respectively. 
Then, 
we analyze the signal gain at UAV with various IRS parameters including the number of elements $K$, 
height of IRS $H_{IRS}$, 
and distance between BS and IRS $L$, 
as well as the height of UAV $H_{UAV}$.
Table~\ref{tab:parameters} summarizes the adopted parameters and the default values used in simulation. 
The results are obtained by averaging 10 thousand simulation runs.

\subsection{Signal Gain versus Number of IRS Elements}
\begin{figure*}[t]
	\centering
	\centering
    \subfigure[]{
		\includegraphics[scale=0.63]{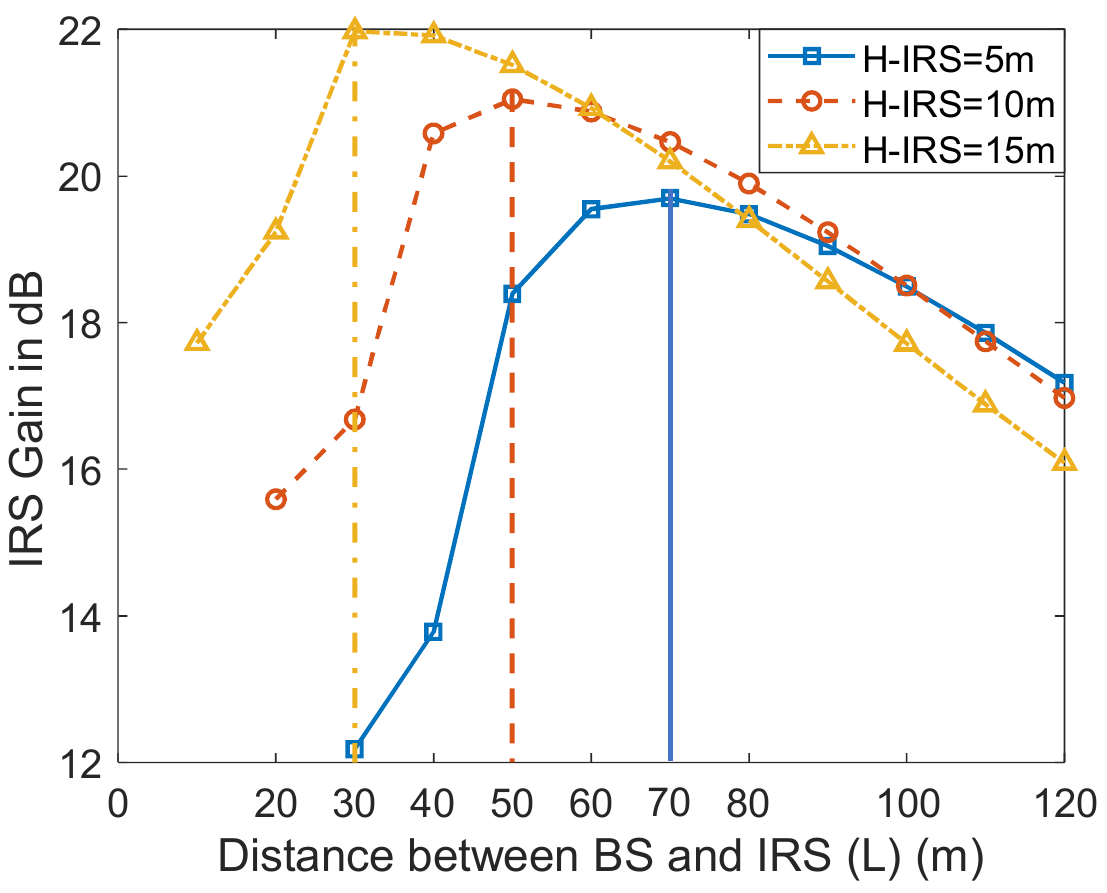}}
		\hspace{0.5in}
	\subfigure[]{
		\includegraphics[scale=0.33]{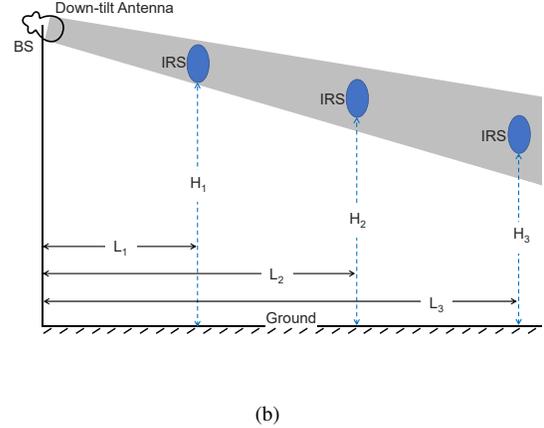}}
	\caption{(a) Impact of IRS height on the seletion of optimal $L$, (b) illustration of the downtilt antenna radiation pattern and its impact on the selection of $L$ and $H_{IRS}$.}
	\label{fig:select_l}
\end{figure*}

Assuming a square IRS, 
we analyze the IRS gain with the number of IRS elements increasing from 25 to 100. 
As plotted in Figure~\ref{fig:irs_numer}, 
the IRS gain increases with the increasing of IRS elements for all the three UAV heights.
Compared to the non-IRS scenario, 
IRS can achieve 21dB received signal gain at UAV when $H_{UAV}$ is 50m. 
When the UAV flies lower, 
the LoS link is within the main lobe of the down-tilt antenna, 
which reduces the IRS gains. 
More interestingly, 
when doubling the number of IRS elements (e.g., from 50 to 100), 
we can observe a 6dB signal gain enhancement. Such observation suggests that the IRS gain scales with the number of IRS elements $K$ in the order of $K^2$. 
As explained in~\cite{wu2018intelligent}, 
the reason arises from two aspects. 
First, 
when increasing $K$, 
the IRS can gather more energy from the BS and reflect more EM waves to the UAV, 
resulting an array gain of $K$. 
Second, 
the capability to reconfigure the phase shift of the reflected EM waves enables a constructive combination of signals at the UAV, 
achieving another gain of $K$.

With the wide deployment of 4G/5G cellular networks, we also evaluate the impact of carrier frequency on the achievable signal gain. Specifically, we fix the height of UAV at 50m and run the same experiments with carrier frequency of 2GHz, 4GHz, and 5GHz, respectively. Our results show that the signal gains are the same for all the frequencies, suggesting IRS is applicable to different generations of cellular networks. Next, we fix $K$ at 100 and evaluate the impact of other parameters.

\vspace{-0.05in}
\subsection{IRS Gain versus UAV Height}

Given that a UAV is always flying in the middle point between BS and IRS (i.e., the horizontal distance between BS and UAV is $L/2$), 
we analyze the impact of UAV height. 
Figure~\ref{fig:2d_distance} (a) plots the IRS gain with $H_{UAV}$ ranging from 20m to 150m, 
in which the range from 20m to 30m is magnified.
We can observe that, 
for a given $L$,
the signal power increases rapidly at low altitude range while saturates when $H_{UAV}$ exceeds 24m. 

 To analyze the reason behind the observed phenomenon, 
 in Figure~\ref{fig:2d_distance} (b), 
 we plot the amplitude of signals received from the LoS, wall, and IRS link in logarithmic scale, respectively. 
 Note the signal gain is the difference of the amplitudes in logarithmic scale. 
 We can see there is a constant gap between IRS link and wall link at all UAV heights and the steep slope is actually caused by the sharp drop in signal strength of the LoS link with $H_{UAV}$ lower than 24m.  

Based on the 3GPP TR36.814~\cite{trseries}, 
the vertical antenna pattern has a very narrow main lobe with angular spread of $26^\circ$.
Given the down-tilt angle of $15^\circ$ and the horizontal distance between BS and UAV of 25m,
the UAV will be out of the main lobe coverage once it flies higher than 24m. 
As a result, 
the energy received from the LoS link will drop significantly while IRS still enjoys the down-tilt antenna pattern to achieve maximum gain. 
In addition, 
we found, in Figure~\ref{fig:2d_distance} (a), that the maximum signal gain achieved at $L=50m$ is higher than those obtained at $L=35m$ and $L=70m$, 
which indicates that there should be an optimal $L$. 
Next, we analyze the impact of $L$.

\begin{figure}[t]
	\centering
	\centering
		\includegraphics[scale=0.63]{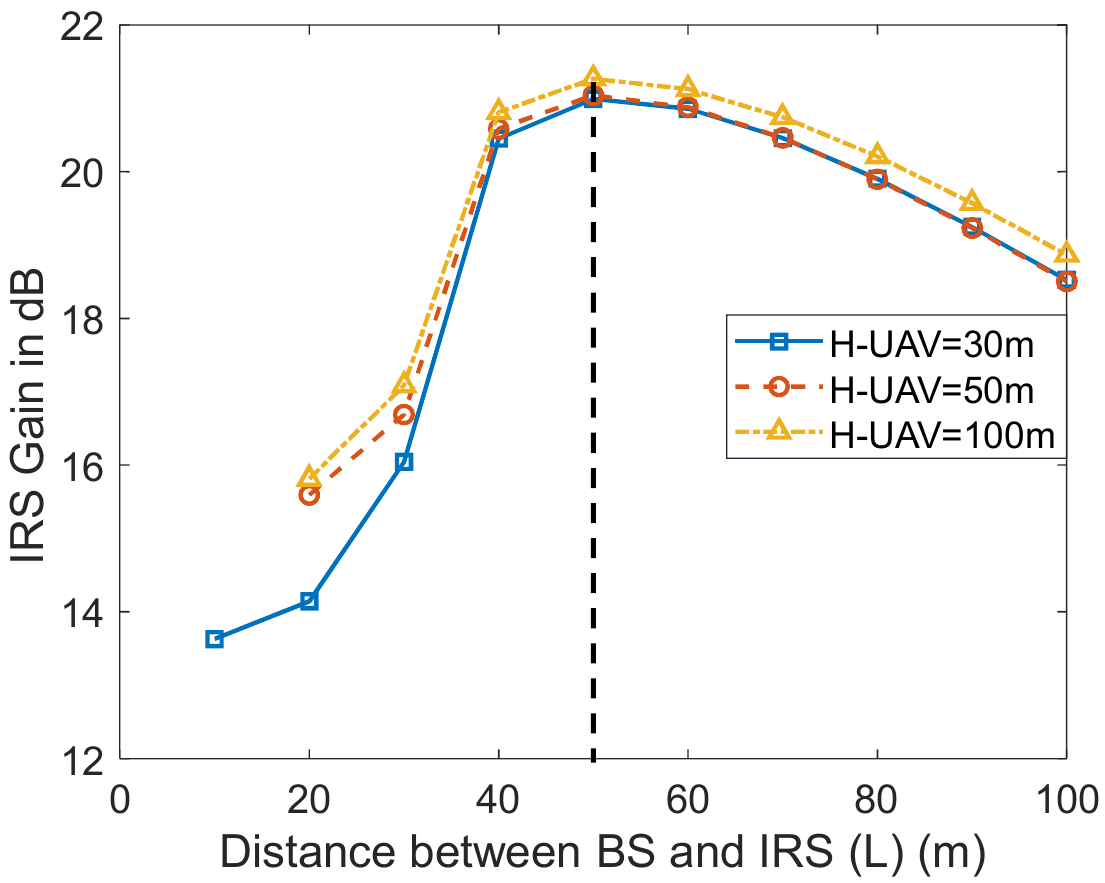}
	\caption{Impact of UAV height on the selection of optimal $L$.}
	\label{fig:uav_height}
% 	\vspace{-0.2in}
\end{figure}

\subsection{Signal Gain versus IRS Location}
Given different height of IRS, Figure~\ref{fig:select_l} (a) plots the signal gain with varying distance $L$ between BS and IRS. 
It can be observed that there is indeed an optimal distance $L$ that yields the highest signal gain, 
although the value varies with $H_{IRS}$. 
For example, 
$L=50m$ is optimal when $H_{IRS}=10m$ while the distance increases to 70m when the IRS is deployed at a lower position ($H_{IRS}=5m$). 
The reason still stems from the down-tilt antenna pattern. 
As shown in Figure~\ref{fig:select_l} (b), 
the gray cone illustrates the coverage of the main lobe of the antenna,
which dominates the radiated power from BS. 
To assist the UAV communication, 
IRS should gather as much energy as possible and re-emit it to the UAV, 
i.e., IRS should be deployed within the gray cone. 
Thus, 
the optimal distance $L$ depends on the height of IRS. 
Such observation is useful to guide the selection of buildings in practical IRS deployment. 

Then, 
we evaluate whether the height of UAV can affect the optimal distance $L$. 
As shown in Figure~\ref{fig:uav_height}, 
we plot the signal gain with different $L$ for when $H_{UAV}$ equals to 30m, 50m, and 100m.
Since $H_{IRS}$ is set to 10m, 
we can see the optimal $L$ is 50m for all the three curves, 
indicating that height of UAV has no impact on the selection of buildings to deploy the IRS.

\section{Related Work}
\label{related}
IRS-assisted wireless communications has emerged as a new field of research with many papers published in the last few years. A majority of the work focused on exploiting IRS as a means for \textit{passive} beamforming to improve the complexity and cost of existing MIMO systems targeting ground users~\cite{huang2018energy,wu2019beamforming,guo2019weighted,yu2019miso,tan2016increasing,wu2018intelligent,he2019cascaded,xu2019resource}. IRS research related to UAV communications is rare with the exception of \cite{li2019reconfigurable}. However, authors of~\cite{li2019reconfigurable} considered UAVs as flying base stations~\cite{fotouhi2018flying} to serve ground users. In our work, we consider the problem of serving UAVs as aerial users through the existing cellular base stations equipped with down-tilted antennas.

\section{Conclusion and Future Work}
\label{s:conclusion}
We have conducted an initial study to assess the potential of employing IRS for improving cellular communications to support UAV users.
Specifically, 
we have considered down-link (BS to UAV) performance for a single-antenna macro BS. 
Our study considered high-precision IRS that can achieve perfect phase and direction control of the EM reflections. 
Based on the 3GPP antenna and path loss models,
we have found that a small IRS patch deployed on building facades can significantly enhance the received cellular signal strengths at the UAVs flying well above the BS antenna, 
where the antenna gain is small due to BS antenna down-tilt. 
We have also found that the gain from IRS is maximized when the IRS deployment altitude and distance from BS are selected optimally. 
This finding provides guidance on IRS deployment to enhance cellular coverage and throughput for flying UAVs.      

This work is an initial study, 
which can be extended in the following several ways: 
\begin{itemize}
\item Our current analysis of UAV mobility is restricted on a 2D plane (i.e., only UAV height and its distance to IRS are considered), while UAVs fly in a 3D space in reality. Thus, it is necessary to investigate the impact of UAV's 3D mobility on the signal gain.

\item Compared to other mobile users (e.g., smartphones), UAVs usually have much higher mobility as they fly fast. As a result, whether IRS-assisted communication can support UAVs under high speed is unclear. If not, how to schedule the movement of UAVs to optimize communication performance should be explored.

\item We assume continuous phase shifts for IRS elements currently and it would be useful to analyze the effect of imperfections in realistic phase controllers that can only switch to some finite phase shifts.

\item Considering IRS deployed on building facades as a common shared resource, we can study IRS resource allocation to support multiple UAVs. In addition, the analysis can be extended to multiple-input-multiple-output (MIMO) scenarios.

\end{itemize}

\balance
\normalem
\bibliographystyle{unsrt} 
\bibliography{ref_ICC}

\end{document}